\documentclass[prl,twocolumn,twoside,showpacs,superscriptaddress]{revtex4}

\usepackage{graphicx}
\usepackage{epsfig}
\usepackage{bm,amsmath,amssymb}
\usepackage{color}

\long\def\comment#1{ }
\newcommand{\eqn}[1]{Eq.~\eqref{#1}}

\newcommand{\beq}{\begin{eqnarray}}
\newcommand{\eeq}{\end{eqnarray}}
\newcommand{\nn}{\nonumber\\}

\newcommand{\rmd}{{\rm d}}
\newcommand{\rme}{{\rm e}}

\newcommand{\del}{\partial}

\newcommand{\bmk}{\bm{k}}

\def\k{{\boldsymbol k}}

\newcommand{\abar}{\bar{\alpha}}

\begin{document}

\title{ Angular structure of the in-medium QCD cascade}

\author{J.-P. Blaizot, Y. Mehtar-Tani and M.A.C. Torres}

\affiliation{Institut de Physique Th\'{e}orique (IPhT) Saclay, CNRS/URA2306,
F-91191 Gif-sur-Yvette, France}

\email{jean-paul.blaizot@cea.fr; yacine.mehtar-tani@cea.fr}

\begin{abstract}

We study the angular broadening of a medium-induced QCD cascade.  We derive the equation that governs the evolution of the average transverse momentum squared of the gluons in the cascade as a function of the medium length, and we solve this equation analytically. Two regimes are identified. For a medium of a not too large size, and for not too soft gluons, the transverse momentum grows with the size of the medium according to standard momentum broadening. The other regime, visible for a medium of a sufficiently large size and very soft gluons, is a regime dominated by multiple branchings: there, the average transverse momentum saturates to a value that is independent of the size of the medium. This structure of the in-medium QCD cascade is, at least qualitatively, compatible with the recent data on dijet asymmetry. 
\end{abstract}
\pacs{12.38.-t, 24.85.+p, 25.75.-q}
\keywords{}
\maketitle

Pb-Pb collisions at the LHC provide rich and new information about the propagation of hard partons in matter. 
 The strong asymmetry of observed dijets \cite{Aad:2010bu,Chatrchyan:2011sx,CMS2014} reflecting the imbalance between the energies of the nearly back--to--back jets, provide direct evidence for the interaction of the jets with the matter produced in the collisions.  A characteristic feature of this interaction is provided by the accurate reconstruction of the missing energy in the sub-leading jet \cite{Chatrchyan:2012ni}, showing that a 
sizable part of this energy is transported by soft particles towards large angles. We  argue in this paper that the latter feature is a genuine property of the in-medium QCD cascades.

Progress in describing such cascades within the framework of perturbative QCD has been made recently. In particular, by assuming that the BDMPSZ  mechanism for medium-induced radiation
\cite{Baier:1996kr,Zakharov:1996fv} dominates the  jet-medium interaction, it was shown \cite{Blaizot:2012fh} that, for a large enough medium, successive gluon emissions  can be considered as independent: multiple emissions can be treated as probabilistic branching processes, with the BDMPSZ spectrum playing the role of
the elementary branching rate \cite{Baier:2000sb,Baier:2001yt,Jeon:2003gi}. The evolution of the energy distribution along such a cascade was studied in \cite{Blaizot:2013hx}, and was shown  to exhibit turbulent behavior. Some features of this cascade where also studied in the context of the thermalization of the quark-gluon plasma \cite{Baier:2000sb,Kurkela:2011ub}. More recently, an equation for the gluon distribution, that takes into account transverse momentum broadening along the cascade was derived in  \cite{Blaizot:2013vha}.  In this paper, we study the first two moments of this equation and provide an analytic analysis of the average angular structure of the in-medium cascade.

Let us recall  that the BDMPS mechanism is characterized by a parameter $\hat q$, called the jet-quenching parameter,  that controls the momentum broadening  in  a direction perpendicular to the jet axis (the average transverse momentum squared acquired by a parton in a medium  is $\langle k_\perp\rangle^2\sim \hat q L$, with $L$ the length of  the medium), as well as the  (radiative) energy loss, in average $\Delta E\sim  \alpha_s\hat q L^2$.  It is valid for frequencies, $\omega_{\rm BH}\lesssim \omega\lesssim \omega_c$, where $\omega_c\sim \hat q L^2$ is the maximum energy that can be taken away by a single 
gluon (the present analysis assumes that $\omega_c\gtrsim E$). 
The lower limit is that of (Bethe-Heitler) incoherent emissions, and is reached when the branching time is of the order of the mean free path between successive collisions. The branching time for a gluon with energy $\omega$ is given by $\tau_{\rm br}(\omega)\sim\sqrt{\omega/\hat q}$. It is associated with a transverse momentum scale $k_{\rm br}\sim(\omega\hat q)^{1/4}$, and an emission angle $\theta_{\rm br}\sim(\hat q/\omega^3)^{1/4}$.  That is, soft gluons are emitted typically at large angles, and on relatively short time scales. To the maximum frequency $\omega_c$, corresponds a minimum emission angle $\theta_c\sim 1/\sqrt{\hat q L^3}$, inside which radiation proceeds essentially as in vacuum.

We focus in this paper on the inclusive gluon distribution function
\beq
D(x,\k,t)= (2\pi)^2x\frac{\rmd N}{\rmd x\rmd^2\k}.
\eeq
Here $x$ is the fraction of the initial energy  $E$ of the leading particle, and $\k$ is the transverse momentum of the gluon observed at time $t$ along the cascade. 
As was shown in \cite{Blaizot:2013vha}, $D(x,\k,t)$ obeys the following evolution equation
\beq\label{Dkt10}
&&\frac{\partial}{\partial t}D(x,\bmk,t)=\frac{1}{4}\nabla_\k^2\left[\hat q\,D(x,\bmk,t)\right]\,\nn
&&+\frac{1}{ t_\ast}\!\int_0^1\!\rmd
z  \,{\cal K}\left(z\right) \!\bigg[\frac{1}{z^2}\sqrt{\frac{z}{x}}
D\left(\frac{x}{z},\k,t\right)-\frac{z}{\sqrt{x}}D\left(x,\bmk,t\right)\bigg], \nn
\eeq
where 
\beq\label{stop-time}
\frac{1}{t_\ast}\equiv\frac{\bar\alpha}{\tau_{_{\rm br}}(E)} = \bar\alpha\sqrt{\frac{\hat q}{E}},\qquad \bar\alpha\equiv \frac{\alpha_s N_c}{\pi},
\eeq
is the basic rate of the branching processes, hence, $t_\ast$ is the typical lifetime of a parton of energy $E$ in the medium. Note that when the size of the medium $t\sim L\gtrsim t_\ast$, or equivalently, for $x$ values such that $xE\lesssim \omega_s\equiv \abar^2\omega_c$, multiple branchings become dominant. The kernel ${\cal K}(z)$  in the gain and loss terms (respectively the second and third term in Eq.~(\ref{Dkt10})), can be written as
\beq\label{Kdef}
{\cal K}(z)=\,\frac{[f(z)]^{5/2}}
{[z(1-z)]^{3/2}}.
\eeq
It collects contributions from  the $z$ dependence of the actual branching time (left out in the definition of $t_\ast$), and from the leading order splitting function $P_{gg}(z)=N_c [f(z)]^2/z(1-z)$ where $N_c$ is the number of colors, and $f(z)\equiv 1-z+z^2$.  (We restrict our discussion to purely gluonic cascades.)
The first term on the r.h.s. of Eq.~(\ref{Dkt10}) is the diffusion term describing transverse momentum broadening. 

Two important approximations are involved in the derivation of Eq.~(\ref{Dkt10}). First, the typical duration of the branching process is assumed to be small compared to the total time spent by the gluon in the medium. This allows to treat  the branchings as effectively instantaneous. Second, the transverse momentum broadening that takes place during a branching is ignored (corrections involving the small transverse momentum induced during the splitting can be absorbed in  corrections to $\hat q$~\cite{Blaizot:2013vha,Blaizot:2014bha}). The branching is then treated as effectively collinear: after the splitting, the two new gluons carry fractions $z$ and $1-z$ of both the initial energy and the initial transverse momentum.   

In order to pin down the main  features of the average angular structure of the in-medium cascade, we study the evolution of the average transverse momentum squared:
\beq\label{kT2}
\langle k_\perp^2\rangle_{t,x}=\frac{\int_{\k} \k^2\, D(x,\k,t)}{\int_{\k}  D(x,\k,t)}\equiv \frac{H(x,t)}{D(x,t)},
\eeq
which involves the first  moments of the gluon distribution. 

The zeroth moment is the energy distribution $D(x,t)=\int_{\k} D(x,\k,t)$. The equation obeyed by $D$ is obtained by integrating \eqn{Dkt10} over $\k$ and reads \cite{Blaizot:2013hx}
 \beq\label{Dfin}
\frac{\partial}{\partial t} D(x,t)\!=\frac{1}{t_\ast}\!\int \rmd z \,
{\cal K}(z)\left[\sqrt{\frac{z}{x}}D\left(\frac{x}{z},t\right)\!-\!\frac{z}{\sqrt{x}}D(x,t)\right].\nn
\eeq
Note that the function $D(x,t)$ has support only for $0\le x\le 1$,
which limits the first $z$-integral in Eq.~(\ref{Dfin}) to $x< z < 1$. Note also that the potential endpoint singularities at $z=1$ in the gain and loss terms cancel, and  \eqn{Dfin} is well defined. The same remark applies to Eq.~(\ref{eqW}) below. 

An analytical solution to this equation exists for the case where, in the kernel (\ref{Kdef}), $f(z)$ is set equal to unity \cite{Blaizot:2013hx}. 
 The simplification of the kernel does not affect its singular behavior  
near $z=0$ and $z=1$, which determines the qualitative features of the solution. We shall use this exact solution from now on. It reads
\beq\label{Dexact}
  D(x,\tau)\,=\,\frac{\tau}{\sqrt{x}(1-x)^{3/2}}\ \rme^{-\pi\frac{\tau^2}
  {1-x}},\qquad \tau\equiv \frac{t}{t_\ast}.\eeq  
 The essential singularity at $x=1$  can be understood as a Sudakov suppression factor  \cite{Baier:2001yt}
 (i.e. the vanishing of the probability to emit no gluon in any finite time).  Aside from this exponential factor, the solution has another remarkable property: for $x\ll 1$,  $ D(x,\tau)\sim (1/\sqrt{x})\, \tau {\rm e}^{-\pi \tau^2}$. The fact that the spectrum keeps the same  $x$-dependence 
when $\tau$ keeps increasing indicates that the energy flows to  $x=0$ without
accumulating  at any finite value of $x$. The complete, energy conserving, solution involves a contribution $\propto\delta(x)$ whose coefficient grows with time as $1-{\rm e}^{-\pi\tau^2}$.  

 Let us now consider the first moment of the distribution, 
$
 H(x,t)= \int_\k \k^2 D\big(x,\k,t\big),
 $
 and set
 $
H(x,t)\equiv x^{2} W(x,t)$ ($W$ is a measure of the square of the angle between the observed gluon and the leading particle).
The equation satisfied by $W(x,t)$ is obtained  by multiplying by $\k^2$ both sides of  Eq.~(\ref{Dkt10}), and integrating over $\k$. One obtains 
   \beq\label{eqW}
  \frac{\del W(x,t)}{\del t}&=& \frac{ \hat q}{x^{2}}\,D(x,t)\nn
  &\!+\!&\!\!\frac{1}{t_\ast}\!\int \rmd z \,{\cal K}(z)\!\left[  \sqrt{\frac{z}{x}} W\Big(\frac{x}{z},t\Big)\!-\!\frac{z}{\sqrt{x}}W\big({x},t\big)  \right]. \nn
  \eeq
 The  initial condition corresponds to the leading particle localized at $x=1$ with no transverse momentum, i.e.,  $W(x,t=0)=0$. Note that 
Eq.~(\ref{eqW}) is identical to  Eq.~(\ref{Dfin}) for $D(x)$, except for the additional  source term  $\sim \hat q/x^2$.  This may be exploited to get the solution  as the following convolution integral
\beq\label{convolW}
W(x,L)=\int_0^t dt \int_x^1\frac{\rmd y}{y} D\left(\frac{x}{y},\frac{L-t}{\sqrt{y}}\right)\,\frac{\hat q}{y^2}\, D(y,t),\nn
\eeq
where $D(x,t)$ is the solution of Eq.~(\ref{Dfin}), written explicitly in Eq.~(\ref{Dexact}). Higher moments can be computed similarly. 

\begin{figure}[!h]
\begin{center}
\includegraphics[width=7.5cm]{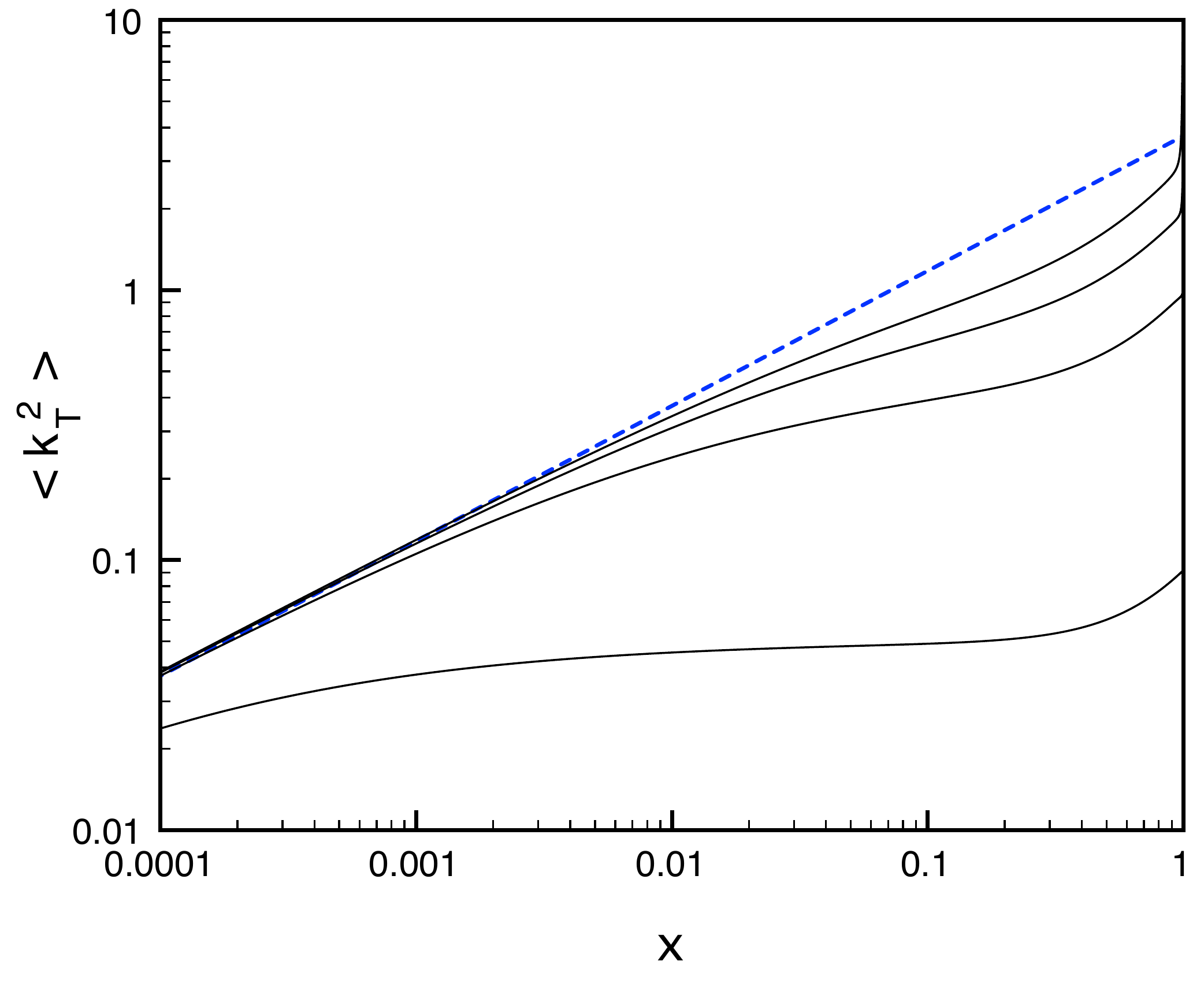}
\caption{The mean transverse momentum squared obtained by solving Eqs.~(\ref{Dkt10}) and (\ref{convolW}), for $t=0.1,1,2,4$ fm (black full lines from the bottom up), for $\hat q=1$ GeV$^2$/fm, $E=100$ GeV and $\bar\alpha=0.3$. The blue dashed line is the asymptotic limit $L\to \infty$ where $\langle k_\perp^2\rangle = (4\bar \alpha)^{-1}\sqrt{xE\hat q }$. Near $x=1$ the curves reproduce the expected behavior, $\langle k_\perp^2\rangle\approx \frac{1}{2}\hat q L\, (1+x^2)$.\label{fig1}}
\end{center}
\end{figure}

The solution of Eq.~(\ref{convolW}) is shown in Fig.~\ref{fig1}. It exhibits two remarkable features:  the average $\langle k_\perp^2\rangle$ decreases  towards small $x$,  $\langle k_\perp^2\rangle$ saturates at a maximum value at large times. These features result from the competition between two effects: the decrease of the transverse momentum that takes place in each branching, the increase of transverse momentum due to collisions. For small times and not too small $x$, momentum broadening due to collisions dominates. But for very small $x$, and/or large time, multiple branchings dominate and are responsible for the saturation of the transverse momentum to a value that depends on $x$ but is independent on time (i.e. on the size of the medium). The transition between the two regimes takes place for  $x\sim x_s=\bar\alpha \hat q L^2/E$. Of course a single regime is visible when $x_s>1$, e.g. for a very large size $L$. 

For short times, and $x>x_s$, one can solve the equation perturbatively, i.e., by iterations. At zeroth order,  by substituting $D^{(0)}(y,t)=\delta(1-y)$ for $D$ in Eq.~(\ref{convolW}), one gets $W^{(0)}(x,L)=\hat q L\delta(1-x)$, so that $\langle k_\perp^2\rangle=\hat q L$: this is the momentum broadening of the leading particle that passes through the whole medium without splitting. The next iteration allows us to examine the effect of a single splitting. This can be calculated easily by using the approximate  expression $D^{(1)}(x,t)\simeq \tau/(\sqrt{x} (1-x)^{3/2})$ that one can read on Eq.~(\ref{Dexact}) for $D(x,t)$. One gets then
$
\langle k_\perp^2\rangle\approx \frac{1}{2}\hat q L\, (1+x^2).
$
It is however instructive to recover this result from a different reasoning. When $x$ is small, but not too small, and the time still short, the leading particle has a chance to split once. Because the splitting is collinear, the transverse momentum squared  carried by the  gluon with frequency $\omega=xE$,   $\langle k_\perp^2\rangle_\omega$, is a fraction of that of the parent gluon, that is,  $\langle k_\perp^2\rangle_\omega=(\omega^2/E^2) \langle k_\perp^2\rangle_E$. Assuming that the splitting occurs at time $t$, the total transverse momentum squared carried by the emitted gluon  is the sum of the transverse momentum inherited from its parent, $x^2\hat q L$, and that it acquires during time $L-t$, $\hat q (L-t)$. Averaging over $t$ yields the result obtained above. Note that  the term proportional to $x^2$, i.e.,  the transverse momentum inherited from the parent gluon before the splitting, becomes negligible at small $x$: all what counts then is the transverse momentum acquired by the observed gluon after it has been emitted.

Let us turn now to the regime dominated by multiple scattering, i.e., $x\ll x_s$ or $L\gg t_\ast$.  The convolution integral in Eq.~(\ref{convolW}) is dominated by small values of $y$, i.e., $1\gg y\gtrsim x$. In this integral, there are two exponential factors, see Eq.~(\ref{Dexact}). The first exponential limits the range of the $t$ integration to  $L-t \lesssim t_\ast \sqrt{x}\ll  t_\ast \sqrt{x_s}\sim L$. The second exponential on the other hand yields $t\lesssim t_\ast<L$. It  follows that in the integration range $[0,L]$, the first function is sharply peaked near $t\sim L$, while $D(y,t)$ is slowly varying. 
One can then set $t\simeq L$ in $D(y,t)$ and integrate $L-t$ from 0 to $\infty$. We then get
\beq
W(x,L)\approx \int_x^1\frac{\rmd y}{y} D(y,L)\frac{\hat q}{y^2}\frac{t_\ast}{2\pi}\frac{\sqrt{y}}{ \sqrt{x/y}\sqrt{1-x/y} },
\eeq
where we have used 
\beq
\int_0^\infty\rmd t\, D(x,t)=\frac{t_\ast}{  2\pi  \sqrt{x(1-x)} }.
\eeq
By noticing that, in the region $x\lesssim y\ll 1$, $D(y,t)/D(x,t)\approx \sqrt{x/y}$, one easily calculates $\langle k_\perp^2\rangle$. One gets
\beq
\langle k_\perp^2\rangle\approx \frac{ \hat q t_\ast\sqrt{x} }{ 2\pi }\int_x^1\rmd u\frac{u}{ \sqrt{u(1-u)} }\approx \frac{k_{\rm br}^2(x)}{4\bar\alpha},
\eeq
where in the last step, we have set the lower bound of the $u$-integral to $0$, and $k_{\rm br}^2(x)=\sqrt{x E \hat q}$. Thus, in the regime dominated by multiple branchings, $\langle k_\perp^2\rangle$ reaches a maximum value that is determined by the local branching transverse momentum $k_{\rm br}^2(x)$, and is independent of the size of the medium. This maximum transverse momentum is also given by $\langle k_\perp^2\rangle\sim\hat q t_\ast(x)$, where $t_\ast(x)=t_\ast\sqrt{x}$ is the lifetime of the observed gluon \cite{Kurkela:2011ub}.

We end the paper by stressing  the phenomenological relevance of the present analysis. 
Assuming that the transverse momentum distribution remains Gaussian in the multiple branching regime, we can postulate the following form for the gluon distribution, 
\beq\label{ansatz}
D(x,\k,L)\simeq D(x,L) \frac{4\pi}{\langle k_\perp^2\rangle}\,\exp\left[-\frac{\k^2}{\langle k_\perp^2\rangle}\right]\,,
\eeq 
 where $\langle k_\perp^2\rangle$ has been determined above as a function of $x$, and  is well approximated by
\beq
\langle k^2_\perp\rangle =\min\left[\frac{1}{2}\hat q t\, (1+x^2), \frac{k^2_\text{br}(x)}{4\bar\alpha},(xE)^2\right].
\eeq
The last condition $k_\perp^2<\omega^2=(xE)^2$ results from the requirement that the emission angle stays smaller than one for our approximations to remain valid. 
One can then estimate  the fraction of the parent gluon (jet) energy that is contained within a cone of size $\Theta$.  This is obtained by integrating the distribution (\ref{ansatz}) over transverse momenta smaller than $\Theta xE$:
\beq\label{in-cone-E}
E_\text{in-cone}(\Theta)=\int_0^1 dx D(x,L) \left[1-\exp\left(-\frac{\Theta^2}{\langle\theta^2\rangle}\right)\right],
\eeq
where $\langle\theta^2\rangle=\langle k_\perp^2\rangle /(xE)^2$. Most of the energy is contained in a small cone of size $\Delta \Theta=\sqrt{\hat qL}/E$ (which in the present analysis is of the same order of magnitude as $\theta_c$). Note that $\Delta\Theta$ is also the angle by which the two jets deviate from being back-to-back.  The maximum energy that can be recovered within a large cone is  $\int_0^1 dx D(x,L)=\exp\left(-\pi L^2/t_\ast^2\right)$. In contrast, an early study, where multiple branchings were not fully taken into account, concluded that the energy carried by medium-induced gluons remains collimated \cite{Salgado:2003rv}. This is less than the initial energy, which reflects a  property of the ideal cascade: because the soft gluons are allowed to split into gluons with arbitrary small frequencies, some energy accumulates at $x=0$, at arbitrarily large angles.  In other words, in   treating the ideal cascade, we have assumed that $\omega_{BH}=0$.  In order to estimate the sensitivity of the results to the value of $\omega_{BH}$, we have solved  Eq. (\ref{in-cone-E}) with a regularized kernel (\ref{Kdef}), viz. $1-z\to 1-(1-x_\text{BH}/x)z$ with $x_\text{BH}=\omega_\text{BH}/E$. With the following set of parameters: $\hat q=1$ GeV$^2$/fm, $\bar\alpha=0.3$,  $\omega_\text{BH}=0, 0.5$ GeV, we find that the leading particle suffers a small broadening, $\Delta \Theta=\sqrt{\hat qL}/E\simeq 0.01-0.02$, after traveling a distance $L=4$ fm.  For $\Theta\ll \Delta \Theta$, the leading particle is out of the cone, so only a small fraction of the energy is in the cone; as soon as $\Theta >  \Delta \Theta$ the leading-particle remains inside the cone where most of the energy ($\sim 70 \%$) is to be found. When the cone angle $\Theta$ increases from 0.1 to 1 the energy barely increases, from 0.7 to 0.75 in the case of the ideal cascade ($\omega_\text{BH}=0$).   The energy is recovered more efficiently for $\omega_\text{BH}=0.5$ GeV: only 10 \% of the total energy  is missing at $\Theta\simeq 1$.
\begin{figure}[!h]
\begin{center}
\includegraphics[width=7.5cm]{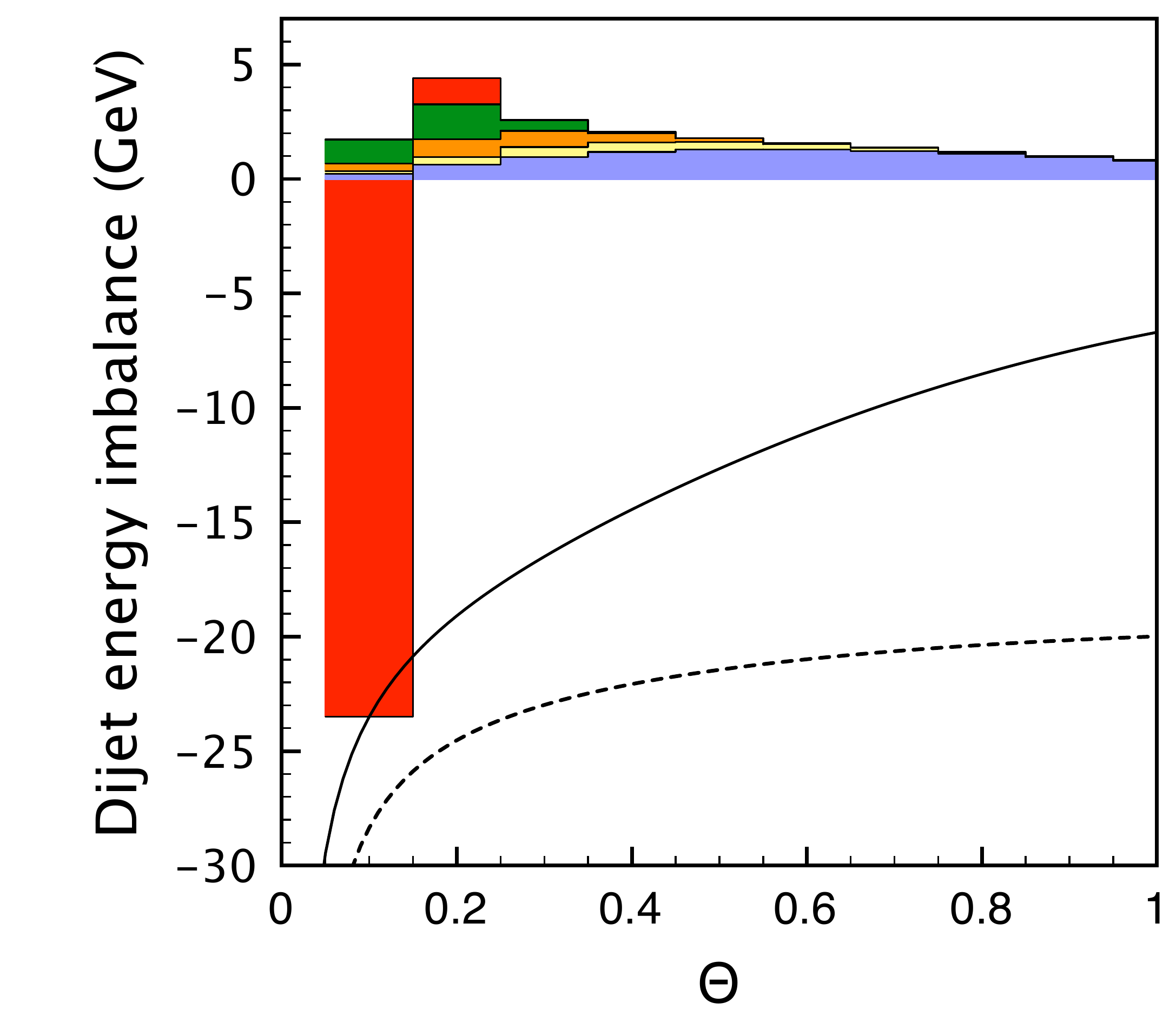}
\caption{The  difference of the angular and energy distributions of sub-leading and leading jets for $\omega_\text{BH}=0.5$ GeV. The histograms account for the four binnings of energies: [0-1] GeV (grey),   [1-2] GeV (yellow), [2-4] GeV(orange), [4-8] GeV(green), [8-100] GeV(red) (color online). In the first angular bin we observe a large imbalance of energy in the hard particles. This energy is partly recovered at large angles by very soft particles.  The cumulative energy is given by the full line. The dashed line represents the limiting case $\omega_\text{BH}=0$, and is given as a reference. The band formed by the full and dashed line can be viewed as a measure of the uncertainty in the treatment of the physics at the scale $\omega_{BH}$.  \label{missing-pt}}
\end{center}
\end{figure}

Finally, we consider the imbalance of energy in dijet events as a function of the energy of the particles and the opening angle of the jet. This observable has been  studied extensively by the CMS collaboration \cite{CMS2014}. Of course, we aim only at a qualitative comparison here.
 We consider the production of two back-to-back jets of 100 GeV each. We assume that the leading jet traverses 1 fm, while  the sub-leading jet traverses 4 fm, in a medium with $\hat q=1$ GeV$^2$/fm. Following CMS, we plot the difference of the energy distribution in the subleading and leading jets in Fig.~\ref{missing-pt}. The asymmetry observed is balanced by soft particles at large angles. The cumulative energy is also given and shows a slow recovery of the energy at large angles for $\omega_\text{BH}=0.5$ GeV. In the limiting case $\omega_\text{BH}=0$, 20 \% of the energy is not balanced due to the uniform flow of energy towards very soft particles and very large angles.

\noindent{\bf Acknowledgements}

This research is supported by the European Research Council under the Advanced Investigator Grant ERC-AD-267258. M.A.C. Torres acknowledges the support from the Brazilian research agency CAPES during his stay at IPhT. 

\providecommand{\href}[2]{#2}\begingroup\raggedright

\endgroup

\end{document}